\documentclass[a4paper, conference]{IEEEtran}

\usepackage{cite}
\usepackage{amsmath,amssymb,amsfonts}
\usepackage{algorithmic}
\usepackage{graphicx}
\usepackage{textcomp}
\usepackage{xcolor}
\usepackage{paralist}
\usepackage{eurosym}
\usepackage{flushend}
\usepackage{url}
\usepackage[utf8]{inputenc}
\usepackage[colorlinks=true]{hyperref}
\usepackage{soul}
\usepackage{caption}
\usepackage[list=true]{subcaption}
\usepackage{booktabs}
\usepackage{tabu}
\usepackage{wrapfig}
\usepackage{arydshln}
\usepackage[normalem]{ulem}
\usepackage{multirow}
\usepackage{siunitx}
\usepackage{adjustbox}
\usepackage{array}
\usepackage{diagbox}
\usepackage{tabularx}
\usepackage{tikz}
\usetikzlibrary{arrows.meta, positioning}
\usepackage{xfrac}
\usetikzlibrary{shapes.geometric, arrows.meta, positioning}
\usetikzlibrary{calc}

\newcommand\blfootnote[1]{%
  \begingroup
  \renewcommand\thefootnote{}\footnote{#1}%
  \addtocounter{footnote}{-1}%
  \endgroup
}

\begin{document}

\title{Active Sensing with Meta-Reinforcement Learning for Emitter Localization from RF Observations}

\def\authorrefmark#1{\ensuremath{^{\textbf{#1}}}}
\author{\IEEEauthorblockN{
M. Shamail J. Khan\authorrefmark{1}, 
Nisha L. Raichur\authorrefmark{1,}\authorrefmark{3}, 
Lucas Heublein\authorrefmark{1}, 
Christian Wielenberg\authorrefmark{1}, 
Alexander Mattick\authorrefmark{1,}\authorrefmark{2}, \\
Tobias Feigl\authorrefmark{1,}\authorrefmark{4},
Christopher Mutschler\authorrefmark{1,}\authorrefmark{2}, 
Felix Ott\authorrefmark{1}}
{\tt\footnotesize\{firstname.lastname\}@iis.fraunhofer.de}\\ 
  \IEEEauthorblockA{\authorrefmark{1}Fraunhofer Institute for Integrated Circuits IIS, 90411 Nürnberg, Germany}
  \IEEEauthorblockA{\authorrefmark{2}Machine Learning and Positioning Systems Lab, University of Technology Nürnberg (UTN), 90461 Nürnberg, Germany}
  \IEEEauthorblockA{\authorrefmark{3}Center for Artificial Intelligence, Technical University of Applied Sciences Würzburg-Schweinfurt, Germany}
  \IEEEauthorblockA{\authorrefmark{4}Programming Systems Lab, Friedrich-Alexander-Universit\"at Erlangen-N\"urnberg, 91052 Erlangen, Germany}
  \vspace{-0.4cm}
}

\maketitle

\begin{abstract}
Global navigation satellite system (GNSS) interference poses a serious threat to reliable positioning, especially in indoor and multipath-rich environments where source localization is highly challenging. In this paper, we formulate GNSS interference localization as an active sensing problem and propose a reinforcement learning (RL) framework in which an agent sequentially explores the environment to infer the position of an emitter source from radio frequency (RF) observations acquired with a $2{\times}2$ patch antenna. The localization task is modeled as a partially observable decision process, since single-snapshot measurements are often ambiguous under multipath propagation and changing channel conditions. To address this, the proposed framework combines high-dimensional RF sensing with deep RL and recurrent policy learning. We investigate both value-based and policy-based approaches, namely Deep Q-Networks (DQN) and Proximal Policy Optimization (PPO), and study their behavior under domain shift. The approach is evaluated on a simulated dataset generated with the Sionna ray-tracing module, which provides realistic propagation effects and diverse environment configurations. Experimental results show that the proposed method achieves a localization success rate of 80.1\%, demonstrating the potential of RL for adaptive GNSS interference localization. Overall, the results highlight simulation-assisted training as a promising direction for robust interference localization in challenging propagation environments.
\end{abstract}
\IEEEpeerreviewmaketitle

\section{Introduction}
\label{label_introduction}

GNSS interference monitoring has advanced considerably in recent years, and methods for interference detection~\cite{borio2016impact} and characterization~\cite{heublein_feigl_jispin} are now well established. Existing approaches can reliably estimate signal properties relevant to mitigation and situational awareness~\cite{borio2021gnss,qin2020assessment,dai2017distortionless,medina2019gnss,bai2020using,lyu2019urban}. The focus is increasingly shifting from detection toward direction finding and localization of the emitter source itself~\cite{schmidt1986multiple,morales2019jammer,heublein_wielenberg}. In this context, localization can be further enhanced by mobile or agent-based sensing platforms that actively explore the environment and collect informative observations, rather than relying on a single static measurement position.

RL provides a natural framework for such sequential decision-making problems~\cite{finn2017model,hansen2021generalization,tang2025deep,li2025deep}. Instead of learning from static input--output pairs, an RL agent interacts with an environment, takes actions, and improves its policy based on the rewards it receives over time~\cite{zhou2023anti,li2019performance,han2017two,zhou2024frequency}. RL has been widely used in control, robotics, and autonomous navigation, where decisions must be made under uncertainty and long-term objectives are important. Recent advances~\cite{li2019performance} in deep RL have significantly expanded the range of solvable problems by combining RL with deep neural networks, enabling learning directly from high-dimensional sensory inputs. Among the most influential methods are value-based approaches such as Deep Q-Networks (DQN)~\cite{van2019jam}, which learn action-value functions from data, and policy-based or actor--critic methods such as Proximal Policy Optimization (PPO)~\cite{schulman2017proximal}, which directly optimize policies while maintaining stable updates.

In this work, we formulate interference localization as an active sensing problem in which a mobile agent sequentially explores the environment to infer the position of an interference source from RF observations measured with an antenna patch. The agent receives local multi-antenna RF observations and selects movement actions over a discrete spatial action space. Rather than treating localization as a one-shot estimation task, the agent is trained to select sensing actions that gradually reduce spatial ambiguity and improve source localization over time. This naturally leads to a partially observable RL problem, since single RF snapshots are often ambiguous in multipath-rich environments and do not uniquely determine the source position. To address this challenge, we investigate deep RL methods (i.e., DQN and PPO), and study how adaptive sensing policies can learn source-seeking behavior under domain shift across different propagation conditions.

\textbf{Contributions.} The main contributions are: (1) We formulate GNSS interference localization as an active sensing problem in which a mobile agent sequentially explores the environment to infer the source position from RF observations. (2) We develop a deep RL framework based on DQN and PPO that learns source-seeking policies directly from high-dimensional measurements acquired with a $2\times2$ patch antenna. (3) We model the task as a partially observable decision process and investigate recurrent policy learning to exploit temporal context under multipath-induced ambiguity. (4) The proposed approach successfully localizes emitter sources in the simulated environment generated with the Sionna ray-tracing framework, achieving a success rate of up to 80.1\%.
\section{Related Work}
\label{label_related_work}

GNSS interference mitigation has traditionally relied on classical signal processing, including frequency-domain analysis~\cite{borio2021gnss}, adaptive filtering~\cite{qin2020assessment}, statistical anomaly detection~\cite{borio2016impact}, suppression~\cite{dai2017distortionless}, and subspace-based direction finding such as MUSIC~\cite{schmidt1986multiple}. While these methods are effective under stable and well-calibrated conditions~\cite{medina2019gnss}, their performance often degrades in realistic environments due to waveform assumptions, calibration errors, dynamic noise, and multipath propagation~\cite{bai2020using}. These limitations motivate data-driven approaches that can learn more robust representations and adapt to varying interference conditions. Supervised approaches such as Random Forests, Decision Trees, and SVMs have shown promising results in distinguishing jamming types and inferring jammer location~\cite{morales2019jammer}. However, these methods often rely on handcrafted features and large labeled datasets, and their performance degrades when interference or propagation conditions differ from training~\cite{lyu2019urban}. Moreover, most existing work focuses on classification rather than active localization, motivating adaptive learning-based methods for real-time emitter source localization.

RL has emerged as a promising approach for dynamic anti-jamming tasks, since it enables agents to learn adaptive policies through interaction with the environment without requiring labeled data. Prior work has shown that (deep) RL methods can outperform rule-based strategies, particularly for channel selection and spectrum-aware decision-making under unknown or intelligent jammers~\cite{li2019performance,van2019jam}. More recent studies have further explored cooperative multi-agent learning, adversarial robustness, and imitation learning to improve adaptation in rapidly changing RF environments~\cite{zhou2023anti}. However, most existing RL-based anti-jamming research focuses on interference avoidance in simplified simulations and low-dimensional state spaces~\cite{li2019performance,van2019jam,han2017two}, rather than source localization from high-dimensional multi-antenna IQ observations~\cite{zhou2024frequency}. These limitations motivate the use of more advanced deep RL frameworks that incorporate temporal modeling and adaptation mechanisms to address partial observability and domain shift~\cite{wang_gong} in emitter localization.

Meta-learning and domain randomization strategies improve the generalization of learning algorithms in non-stationary environments. Meta-learning enables rapid adaptation to new tasks from limited data by learning model parameters that can be efficiently fine-tuned, as demonstrated by methods such as MAML~\cite{finn2017model}. Domain randomization improves robustness by training agents across a distribution of environment variations, thereby reducing sensitivity to specific training conditions~\cite{hansen2021generalization}. In RF learning, these ideas are especially relevant because changes in propagation conditions, interference patterns, and physical layouts can induce substantial domain shift that cannot be resolved by simple signal-level augmentation alone~\cite{tang2025deep,li2025deep}. Although prior work has shown the value of meta-learning and randomized training for robust RL, their use for emitter localization in multipath-rich GNSS environments remains largely unexplored.

\section{Preliminaries}
\label{label_preliminaries}

\subsection{Markov Decision Process (MDP)}

RL problems are commonly formulated as an MDP by
\begin{equation}
    \mathcal{M} = \langle \mathcal{S}, \mathcal{A}, \mathcal{P}, \mathcal{R}, \gamma \rangle,
\end{equation}
where $\mathcal{S}$ denotes the state space, $\mathcal{A}$ the action space, $\mathcal{P}$ the transition model, $\mathcal{R}$ the reward function, and $\gamma \in [0,1]$ the discount factor~\cite{puterman2014markov}. MDP assumes the Markov property, i.e., the next state depends only on the current state and action:
\begin{equation}
\resizebox{0.91\linewidth}{!}{$
    P(s_{t+1}\mid s_t,a_t) = P(s_{t+1}\mid s_t,a_t,s_{t-1},a_{t-1},\dots,s_0,a_0).
$}
\end{equation}
A policy $\pi(a\!\mid\!s)$ specifies the probability of selecting action $a$ in state $s$~\cite{sutton1998reinforcement}. The objective is to maximize the expected discounted return $G_t = \sum_{k=0}^{\infty} \gamma^k r_{t+k+1}$. The corresponding state-value and action-value functions are given by
\begin{equation}
\resizebox{0.91\linewidth}{!}{$
    V^{\pi}(s) = \mathbb{E}_{\pi}[G_t\!\mid\!S_t=s], \,\,\, Q^{\pi}(s,a) = \mathbb{E}_{\pi}[G_t\!\mid\!S_t=s, A_t=a].
$}
\end{equation}
The optimal action-value function satisfies the Bellman optimality equation:
\begin{equation}
\resizebox{0.91\linewidth}{!}{$
    Q^{\ast}(s,a) = \mathbb{E}\!\left[r_{t+1} + \gamma \max_{a'} Q^{\ast}(s_{t+1},a')\!\mid\!S_t=s, A_t=a \right].
$}
\end{equation}
We model the problem as a partially observable MDP (POMDP), $\langle \mathcal{S}, \mathcal{A}, \mathcal{P}, \mathcal{R}, \Omega, \mathcal{O}, \gamma \rangle$~\cite{kaelbling1998planning}, where $\Omega$ denotes the observation space and $\mathcal{O}(o\!\mid\!s,a)$ the observation model~\cite{kaelbling1998planning}. Since individual RF observations are often ambiguous, the policy must exploit temporal context.

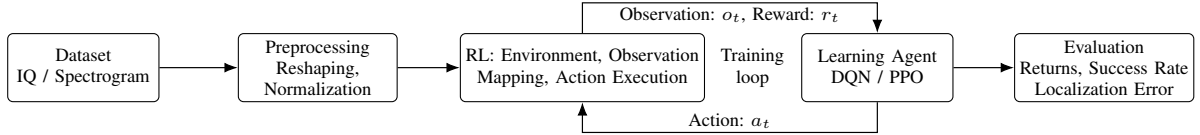
\begin{figure*}[t]
    \centering
    \scriptsize
    \begin{tikzpicture}[
        >=Latex,
        every node/.style={align=center, font=\scriptsize},
        box/.style={
            draw,
            rounded corners=2pt,
            minimum height=0.9cm,
            minimum width=2.0cm,
            inner sep=2pt
        },
        flow/.style={->, line width=0.45pt}
    ]

    \node[box] (data) at (0,0) {Dataset\\IQ / Spectrogram};

    \node[box, minimum width=2.1cm] (prep) at (3.1,0)
    {Preprocessing\\Reshaping,\\Normalization};

    \node[box, minimum width=2.2cm] (env) at (6.6,0)
    {RL: Environment, Observation\\Mapping, Action Execution};

    \node[box, minimum width=2.0cm] (agent) at (10.5,0)
    {Learning Agent\\DQN / PPO};

    \node[box, minimum width=2.2cm] (eval) at (13.5,0)
    {Evaluation\\Returns, Success Rate\\Localization Error};

    \draw[flow] (data.east) -- (prep.west);
    \draw[flow] (prep.east) -- (env.west);
    \draw[flow] (agent.east) -- (eval.west);

    \draw[flow] (env.north) -- ++(0,0.4) -- ++(3.9,0) -- (agent.north);

    \draw[flow] (agent.south) -- ++(0,-0.4) -- ++(-3.9,0) -- (env.south);

    \node[font=\scriptsize] at (8.55,0.7) {Observation: $o_t$, Reward: $r_t$};
    \node[font=\scriptsize] at (8.55,-0.7) {Action: $a_t$};
    \node[font=\scriptsize] at (8.85,0) {Training\\loop};

    \end{tikzpicture}
    \vspace{-0.2cm}
    \caption{End-to-end pipeline showing preprocessing, environment interaction, agent training, and evaluation.}
    \label{figure_pipeline}
\end{figure*}

\subsection{Deep Reinforcement Learning (DRL)}

DRL uses deep neural networks as function approximators for value functions and policies~\cite{mnih2013playing}. This enables learning from high-dimensional observations, where tabular methods become infeasible. We consider two representative DRL algorithms: value-based approaches, such as DQN, and policy-based approaches, such as PPO, with actor--critic methods combining both formulations.

\paragraph{Deep Q-Networks} DQN approximates the optimal action-value function $Q^{\ast}(s,a)$ with a neural network $Q(s,a;\theta)$~\cite{mnih2013playing}. To stabilize training, DQN combines experience replay with a target network, yielding the loss
\begin{equation}
\resizebox{1.0\linewidth}{!}{$
    L_i(\theta_i)=\mathbb{E}_{(s,a,r,s')\sim U(D)}
    \left[
    \left(
    r+\gamma \max_{a'} Q(s',a';\theta_i^-)-Q(s,a;\theta_i)
    \right)^2
    \right],
$}
\end{equation}
where $\theta_i^-$ denotes the parameters of the target network. DQN is effective for discrete decision-making problems, but its reliance on single observations can limit performance in partially observable environments.

\paragraph{Policy-Gradient} PPO is designed to improve the stability of policy optimization by constraining the update between successive policies~\cite{schulman2017proximal}. It maximizes the clipped surrogate objective
\begin{equation}
\resizebox{0.91\linewidth}{!}{$
    L^{\mathrm{CLIP}}(\theta)=
    \hat{\mathbb{E}}_t
    \left[
    \min\!\left(
    r_t(\theta)\hat{A}_t,\,
    \mathrm{clip}\!\left(r_t(\theta),1-\epsilon,1+\epsilon\right)\hat{A}_t
    \right)
    \right],
$}
\end{equation}
where
\begin{equation}
    r_t(\theta) = \frac{\pi_{\theta}(a_t\mid s_t)}{\pi_{\theta_{\mathrm{old}}}(a_t\mid s_t)}
\end{equation}
is the probability ratio and $\hat{A}_t$ denotes the estimated advantage. PPO is commonly implemented in an actor--critic architecture, where the actor represents the policy and the critic estimates the state value. Owing to its stable updates, PPO is well suited to noisy and partially observable problems.

\paragraph{Recurrent Architectures} In partially observable settings, the current observation may not uniquely identify the underlying state. Recurrent architectures address this limitation by integrating temporal context over multiple time steps. In particular, long short-term memory (LSTM) networks maintain an internal memory state through gated updates. By aggregating information over time, recurrent DRL agents can better resolve ambiguity in sequential RF observations and improve state estimation under multipath and domain shift.


\section{Methodology}
\label{label_method}

\subsection{Problem Formulation \& Key Idea}
\label{label_method_problem_formulation}

\begin{figure}[!b]
    \centering
    \begin{minipage}[t]{0.63\linewidth}
        \centering
        \vspace{0pt}
        \includegraphics[trim=26 34 28 44, clip, width=1.0\linewidth]{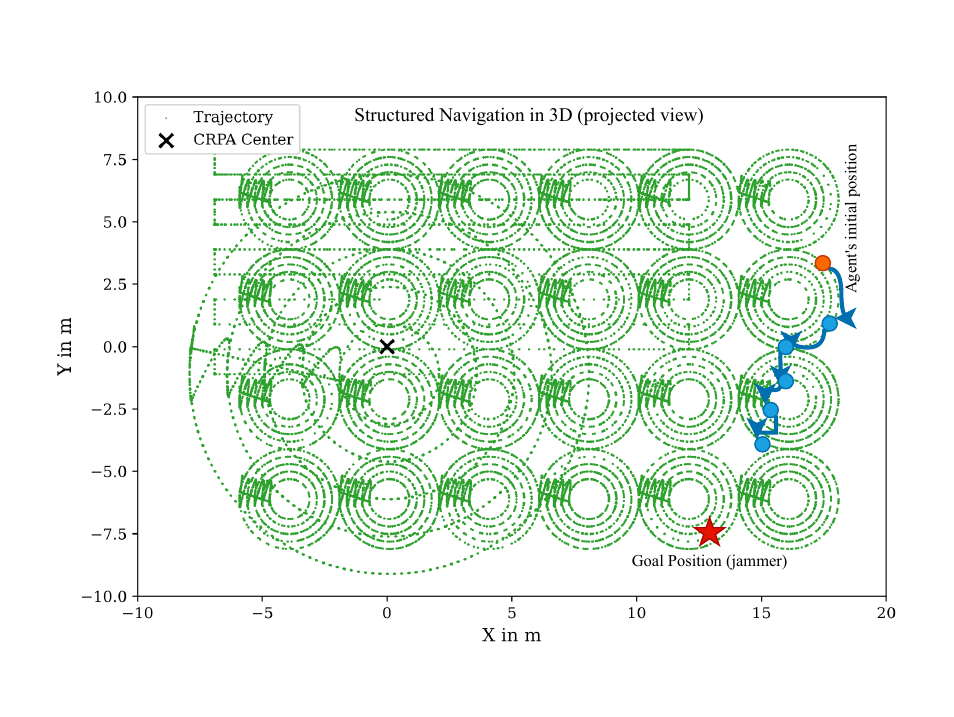}        
        \caption{Structured navigation environment of the agent.}
        \label{figure_environment_agent}
    \end{minipage}
    \hfill
    \begin{minipage}[t]{0.35\linewidth}
    \centering
    \vspace{0pt}
    \begin{tikzpicture}[
        scale=0.42,
        transform shape,
        >=Latex,
        every node/.style={font=\footnotesize},
        flow/.style={line width=0.65pt}
    ]

    \draw[
        black,
        line width=2.2pt,
        line cap=round,
        smooth,
        variable=\t,
        domain=0:18,
        samples=140
    ]
    plot ({(2.35 - 0.10*\t)*cos(\t r)},
          {(2.35 - 0.10*\t)*sin(\t r)});

    \draw[flow, red!85!black, ->]
        ($(150:2.75)$) arc[start angle=150,end angle=210,radius=2.75];
    \draw[flow, red!85!black, ->]
        ($(142:3.10)$) arc[start angle=142,end angle=210,radius=3.10];

    \draw[flow, blue!85!black, ->]
        ($(38:2.75)$) arc[start angle=38,end angle=-5,radius=2.75];
    \draw[flow, blue!85!black, ->]
        ($(52:3.10)$) arc[start angle=52,end angle=-5,radius=3.10];

    \draw[flow, green!85!black, <->]
        ($(225:0.6)$) -- ($(225:2.00)$);

    \node[text=red!85!black]  at (-1.2,2.6) {\large 2 Counter Clockwise Actions};
    \node[text=blue!85!black] at (2.0,3.0) {\large 2 Clockwise Actions};
    \node[text=green!85!black, rotate=-30] at (-1.7,-1.8) {\large Radial};

    \end{tikzpicture}
    \caption{Discrete action space for structured navigation.}
    \label{figure_action_space}
    \end{minipage}
\end{figure}

We consider the problem of localizing a stationary GNSS interference source from RF observations acquired with a $2 \times 2$ antenna patch. Rather than treating localization as a single-step estimation problem, we formulate it as an active sensing task without grid-based localization in which a mobile agent sequentially selects sensing locations to reduce uncertainty about the source position, see Figure~\ref{figure_environment_agent}. This view is motivated by the fact that a single RF snapshot is generally insufficient to determine the 3D source location, especially in indoor environments where multipath propagation and non-line-of-sight components introduce strong ambiguity. By moving through the environment and observing how the received signals change across space, the agent can progressively refine its estimate of the source location. The overall pipeline of preprocessing, environment interaction, policy learning, and evaluation, see Figure~\ref{figure_pipeline}.

The localization task is modeled as a partially observable sequential decision process. At timestep $t$, the underlying state is defined as $s_t=(p_t,p_I)$, where $p_t \in \mathbb{R}^3$ denotes the agent position and $p_I \in \mathbb{R}^3$ denotes the fixed but unknown interference source position. Given an action $a_t$ (see Figure~\ref{figure_action_space}, for the action space), the agent moves according to the environment dynamics to a new position $p_{t+1}$. In contrast, the observation process is stochastic: at each position, the agent receives an RF observation $o_t$ influenced by propagation effects, noise, and reflections. Consequently, similar observations may arise at distinct locations, such that the source position cannot be inferred reliably from a single measurement.

At each timestep, the agent selects an action $a_t \in \mathcal{A}$, receives the next observation $o_{t+1}$, and obtains a reward $r_t$. The objective is to learn a policy $\pi(a_t\!\mid\!o_t)$ that maximizes the expected cumulative discounted reward
\begin{equation}
\label{eq:objective_function}
\resizebox{0.37\linewidth}{!}{$
    J(\pi) = \mathbb{E}_{\pi}\!\left[\sum_{t=0}^{T} \gamma^t r_t \right],
$}
\end{equation}
while reaching the source region defined by $\|p_t-p_I\|_2 \leq \epsilon$, where $\epsilon$ is a predefined success threshold~\cite{sutton1998reinforcement}. Since the observations are only partially informative, effective localization requires integrating information across time. In this work, this is achieved by RL policies that exploit observation histories to guide source-seeking behavior.

\subsection{Reward Function Design}
\label{label_method_reward_function_design}

The reward function is designed to encourage efficient source-seeking behavior under partial observability. Since the interference source position is not directly observable, the reward must provide informative feedback based on spatial progress rather than instantaneous signal features, which may be unreliable under noise and multipath~\cite{kaelbling1998planning}. We therefore adopt a simple reward design based on progress, step cost, and terminal success. The main component is a distance-based progress reward. Let
\begin{equation}
    d_t = \|p_t - p_I\|_2
\end{equation}
denote the Euclidean distance between the agent position $p_t$ and the interference source position $p_I$ at timestep $t$. Following potential-based reward shaping~\cite{Ng1999PolicyIU}, the progress reward is defined as
\begin{equation}
    r_{\text{prog}}(t) = \alpha (d_{t-1} - d_t),
\end{equation}
where $\alpha > 0$ is a scaling factor. This term is positive when the agent moves toward the source and negative otherwise. To promote efficient trajectories, we additionally impose a constant step penalty $r_{\text{step}} = -\lambda$, with $\lambda > 0$. The agent receives a terminal reward upon entering the success region,
\begin{equation}
    r_{\text{succ}} =
    \begin{cases}
        R_{\text{goal}}, & \text{if } \|p_t - p_I\|_2 \leq \epsilon,\\
        0, & \text{otherwise},
    \end{cases}
\end{equation}
where $\epsilon$ denotes the localization threshold and $R_{\mathrm{goal}}$ is a positive constant. The total reward at timestep $t$ is given by
\begin{equation}
    r_t = r_{\text{prog}}(t) + r_{\text{step}} + r_{\text{succ}}.
    \label{eq:total_reward}
\end{equation}

In preliminary experiments, more elaborate shaping terms led to unstable training and increased return variance. So, we retain the above reward design for its simplicity and stability for generalization across different propagation environments.

\subsection{Policy \& Value Function Architecture}
\label{label_method_policy_value_function_architecture}

\begin{figure}[t]
    \centering
    \scriptsize
    \begin{tikzpicture}[
        >=Latex,
        node distance=0.55cm,
        every node/.style={align=center, font=\scriptsize},
        box/.style={
            draw,
            rounded corners=2pt,
            minimum height=0.95cm,
            minimum width=2.2cm,
            inner sep=2pt
        },
        smallbox/.style={
            draw,
            rounded corners=2pt,
            minimum height=0.6cm,
            minimum width=1.1cm,
            inner sep=2pt
        },
        flow/.style={->, line width=0.5pt}
    ]
    
    \node[box] (input) {IQ Statistics input\\$\mathbf{X} \in \mathbb{R}^{N \times 2 \times 4 \times d}$};
    \node[box, right=0.7cm of input] (enc) {1D-CNN encoder\\$(8,1024)\rightarrow(256)$};
    \node[smallbox, right=0.7cm of enc] (flat) {Flatten\\$(256)$};
    \node[smallbox, right=0.7cm of flat, yshift=0.55cm] (actor) {Actor\\$\pi$};
    \node[smallbox, right=0.7cm of flat, yshift=-0.55cm] (critic) {Critic\\$V$};
    
    \draw[flow] (input) -- (enc);
    \draw[flow] (enc) -- (flat);
    
    \draw[flow] (flat.east) -- ++(0.28,0) |- (actor.west);
    \draw[flow] (flat.east) -- ++(0.28,0) |- (critic.west);
    \end{tikzpicture}
    \vspace{-0.5cm}
    \caption{Feedforward PPO architecture.}
    \label{figure_memoryless_ppo}
\end{figure}
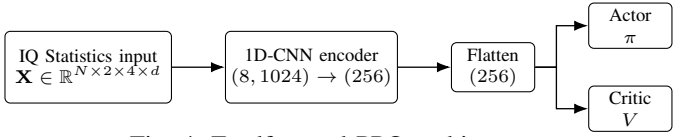

The model input is represented as a structured tensor $\mathbf{X} \in \mathbb{R}^{N \times 2 \times 4 \times d}$, where $N$ denotes the batch size, the second dimension distinguishes between the current and goal states, the third dimension corresponds to the four antenna elements, and $d$ denotes the number of features per antenna. Accordingly, each entry $\mathbf{X}_{n,t,i} \in \mathbb{R}^{d}$ represents the feature vector associated with antenna $i \in \{1,2,3,4\}$ for state type $t \in \{\text{current}, \text{goal}\}$.

For each sample, the feature vectors of the four antennas associated with the current and goal states can, if required by the downstream network, be concatenated into a single input vector, yielding a representation in $\mathbb{R}^{8d}$. This formulation preserves the individual contribution of each antenna while jointly encoding the current and target states within a unified representation. The proposed representation preserves the full feature resolution of each antenna and enables subsequent layers to learn both inter-antenna dependencies and the relationships between the current and goal configurations. To process this high-dimensional input efficiently, we employ a shared feature extractor implemented as a 1D convolutional encoder, which maps the structured observation to a compact latent representation. The encoder is designed to capture both local structure in the feature space and dependencies across antenna channels.

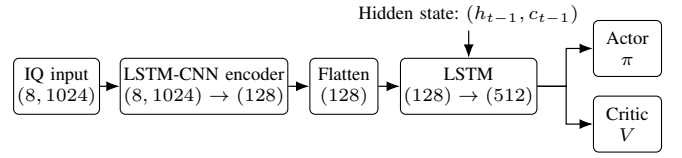
\begin{figure}[t]
    \centering
    \scriptsize
    \begin{tikzpicture}[
        >=Latex,
        every node/.style={align=center, font=\scriptsize},
        box/.style={
            draw,
            rounded corners=2pt,
            minimum height=0.74cm,
            minimum width=1.45cm,
            inner sep=1.2pt
        },
        smallbox/.style={
            draw,
            rounded corners=2pt,
            minimum height=0.74cm,
            minimum width=0.9cm,
            inner sep=1.2pt
        },
        flow/.style={->, line width=0.45pt}
    ]

    \node[box, minimum width=1.1cm] (input) at (0,0) {IQ input\\$(8,1024)$};
    \node[box, minimum width=1.75cm] (enc)   at (1.95,0) {LSTM-CNN encoder\\$(8,1024)\rightarrow(128)$};
    \node[box, minimum width=0.9cm] (flat) at (3.8,0) {Flatten\\$(128)$};
    \node[box, minimum width=1.45cm] (lstm)  at (5.45,0) {LSTM\\$(128)\rightarrow(512)$};

    \node[smallbox] (actor)  at (7.55,0.5) {Actor\\$\pi$};
    \node[smallbox] (critic) at (7.55,-0.5) {Critic\\$V$};

    \node[font=\scriptsize] (hidden) at (5.45,0.95) {Hidden state: $(h_{t-1},c_{t-1})$};

    \draw[flow] (input.east) -- (enc.west);
    \draw[flow] (enc.east) -- (flat.west);
    \draw[flow] (flat.east) -- (lstm.west);

    \draw[flow] (lstm.east) -- ++(0.38,0) |- (actor.west);
    \draw[flow] (lstm.east) -- ++(0.38,0) |- (critic.west);

    \draw[flow] (hidden.south) -- (lstm.north);

    \end{tikzpicture}
    \vspace{-0.1cm}
    \caption{Recurrent PPO architecture.}
    \label{figure_recurrent_ppo}
\end{figure}

Since interference localization is modeled as a partially observable decision process, a single RF snapshot is generally insufficient to uniquely infer the source position. We therefore consider both feedforward and recurrent architectures. The feedforward model consists of the convolutional encoder followed directly by policy and value heads (Figure~\ref{figure_memoryless_ppo}). The recurrent model extends this architecture by incorporating an LSTM module that aggregates information over time (Figure~\ref{figure_recurrent_ppo}). This temporal memory enables the agent to exploit sequential signal variations and to construct an implicit belief state in the presence of multipath-induced ambiguity.

The latent representation is shared between the actor and critic networks in order to improve parameter efficiency and enhance training stability. The actor produces a categorical distribution over the discrete action space, while the critic estimates the corresponding state value. This actor--critic architecture is employed in conjunction with PPO to enable stable policy optimization in noisy environments. In addition, DQN is considered as a value-based baseline using the same feature encoder and, for the recurrent variant, the same temporal module. Overall, the proposed architecture combines convolutional feature extraction, optional recurrent memory, and shared policy--value learning to support adaptive source localization under uncertainty.

\begin{table}[t]
\setlength{\tabcolsep}{1.8pt}
\centering
\footnotesize
\caption{Training hyperparameters.}
\label{tab:hyperparameters}
\begin{tabular}{l p{0.74\linewidth}}
    \toprule
    \textbf{Method} & \textbf{Settings} \\
    \midrule
    DQN & $10^6$ steps, Adam ($\epsilon{=}10^{-5}$), $\gamma{=}0.99$, lr $=2.5{\cdot}10^{-4}$, TU $=1000$, $\epsilon{:}1.0\!\rightarrow\!0.05$ over $20\%$, update every $4$ steps. FF: RB $=50$k, BS $=128$, learn start $=10$k. Rec: RB $=5$k, BS $=32$, learn start $=5$k, seq len $=8$. \\
    PPO & $10^6$ steps, Adam ($\epsilon{=}10^{-5}$), $8$ envs, lr $=2.5{\cdot}10^{-4}$ (annealed), $\gamma{=}0.997$, GAE $\lambda{=}0.95$, clip $=0.1$, ent $=0.02$, value $=0.5$, grad clip $=0.5$, MB $=128$, epochs $=4$. \\
    ANIL-FOMAML & $2000$ meta-iters, $3$ tasks/batch, outer lr $\beta{=}10^{-4}$, critic lr $=10^{-4}$, grad clip $=0.5$, inner lr $\alpha{=}3{\cdot}10^{-3}$, $1$ adapt step, support $=128$ steps (train) / $5$ episodes (eval), $\gamma{=}0.997$, clip $=0.1$, ent $=0.02$, value $=0.4$. \\
    \bottomrule
\end{tabular}
\vspace{-0.2cm}
\end{table}

\subsection{Statistical Feature Selection for Policy Learning}
\label{label_statistical_feature_selection}

To enable the RL agent to navigate the environment, we extract a set of statistical features from the IQ samples and train the agent to localize the emitter using each feature individually. Specifically, we consider five feature types: the IQ mean, the IQ standard deviation, the IQ root-mean-square (RMS), and the IQ phase differences between antenna pairs. Based on these feature representations, the agent is trained using the policy and value function architecture described in Section~\ref{label_method_policy_value_function_architecture}. In addition, during training, each episode is associated with a randomly assigned goal position. The environment is reset once the minimum number of steps has been reached. Consequently, the agent is not trained to reach a single fixed target, but instead learns a general policy that enables navigation toward arbitrary goal positions within the considered spatial region.

\section{Experiments}
\label{label_experiments}

\paragraph{Sionna Simulation Environment} All experiments are conducted on a synthetic dataset generated with the Sionna ray-tracing framework~\cite{hoydis_cammerer}. The simulated scene represents an indoor industrial hall with metallic boundary surfaces and interior structures such as shelves, yielding dense multipath propagation and non-line-of-sight conditions. Sionna computes the propagation paths and generates the corresponding complex baseband signals, which are then used as input to the localization framework. The transmitter is placed at a fixed position in the scene and operates at the GPS L1 center frequency of $1.57542\,\text{GHz}$. The receiver is modeled as a $2 \times 2$ antenna patch that traverses the environment while maintaining a fixed orientation. For each receiver position, the simulator generates $1{,}024$ complex IQ samples per antenna element at a bandwidth of $100\,\text{MHz}$. Each observation is represented as a tensor of shape $(4 \times 1{,}024 \times 2)$.

\begin{figure}[!t]
    \centering
    \begin{minipage}[t]{0.493\linewidth}
        \centering
        \includegraphics[trim=10 6 6 8, clip, width=1.0\linewidth]{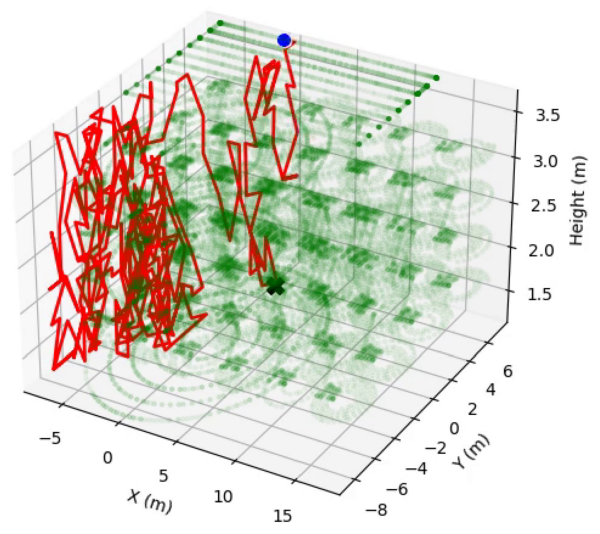}
    \end{minipage}
    \hfill
    \begin{minipage}[t]{0.493\linewidth}
        \centering
        \includegraphics[trim=10 6 6 8, clip, width=1.0\linewidth]{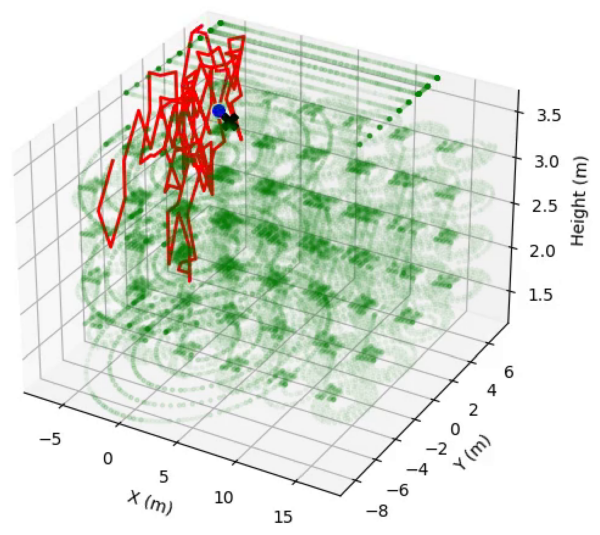}
    \end{minipage}
    \caption{Trajectories of the agent during early (left) and late (right) training phase with PPO.}
    \label{figure_baseline_trajectory}
    \vspace{-0.5cm}
\end{figure}

\paragraph{Signal Structure \& Multipath Propagation} Let $s(t)$ denote the transmitted complex baseband signal. The received signal at the antenna array is modeled as a superposition of multiple propagation paths,
\begin{equation}
    \mathbf{y}(t)=\sum_{l=1}^{L}\alpha_l \mathbf{a}(\theta_l,\phi_l)s(t-\tau_l)+\mathbf{n}(t),
\end{equation}
where $\alpha_l$, $\tau_l$, and $(\theta_l,\phi_l)$ denote the complex attenuation, delay, and direction of arrival of the $l$-th path, respectively, and $\mathbf{a}(\theta_l,\phi_l)$ is the array response. This model reflects the strong multipath conditions of the simulated indoor scene, where reflections from walls, floor, ceiling, and interior objects create ambiguous spatial signatures. Since the set of multipath components depends on the environment geometry, changes in the scene induce domain shifts in the observation distribution.

\paragraph{Data Preprocessing} The proposed framework operates directly on raw IQ observations in order to preserve both amplitude and phase information. Before training, the IQ samples are standardized channel-wise using z-score normalization, $x_{\mathrm{norm}} = \frac{x-\mu}{\sigma}$, where $\mu$ and $\sigma$ are computed on the training set and reused for all evaluation settings. This preprocessing ensures numerical stability and a consistent input scale across environments, while maintaining the signal structure required for learning-based localization.

\section{Evaluation}
\label{label_evaluation}

Overall, the results show that RL can learn meaningful source-seeking behavior from RF observations, while recurrence and meta-learning provide further gains under partial observability and domain shift.

\paragraph{Evaluation Metrics} The \textbf{episodic return} measures the cumulative reward obtained within one episode, $G = \sum_{t=0}^{T} r_t$, and serves as the primary indicator of overall task performance, reflecting navigation efficiency and successful localization. The \textbf{success rate} (SR) is defined as the fraction of evaluation episodes in which the agent reaches the target region, $\mathrm{SR} = \frac{1}{N}\sum_{i=1}^{N}\mathbb{I}[\mathrm{success}_i]$, where $N$ denotes the number of evaluation episodes. For PPO-based methods, the \textbf{explained variance} is $\mathrm{EV} = 1 - \frac{\mathrm{Var}(V_{\mathrm{target}} - V_{\mathrm{pred}})}{\mathrm{Var}(V_{\mathrm{target}})}$, where $V_{\mathrm{target}}$ is the empirical return and $V_{\mathrm{pred}}$ is the critic estimate. Higher EV indicates a more accurate value function and improved state representation. For meta-learning, we evaluate the \textbf{mean query return}, which measures post-adaptation performance on the query set, $J_{\mathrm{query}}(\theta) = \frac{1}{|\mathcal{B}|} \sum_{\mathcal{T}_i \in \mathcal{B}} \mathbb{E}_{\tau \sim \pi_{\theta_i'}} \left[ \sum_{t=0}^{T} \gamma^t r_t \right]$, where $\theta_i'$ denotes the task-adapted parameters.

\paragraph{Baseline Feasibility} As a first baseline, PPO outperformed DQN, achieving a success rate of $80.1\%$ compared to $76.3\%$ and a higher mean return ($4.98$ vs.\ $-0.09$). Although localization errors remained high, these results indicate that RL can extract useful navigation cues from RF observations and learn non-trivial interference-seeking behavior. This qualitative progression is illustrated in Figure~\ref{figure_baseline_trajectory}, where the PPO agent shows more goal-directed trajectories in later training stages.

\begin{figure}[!t]
    \centering
        \includegraphics[trim=6 6 6 20, clip, width=0.65\linewidth]{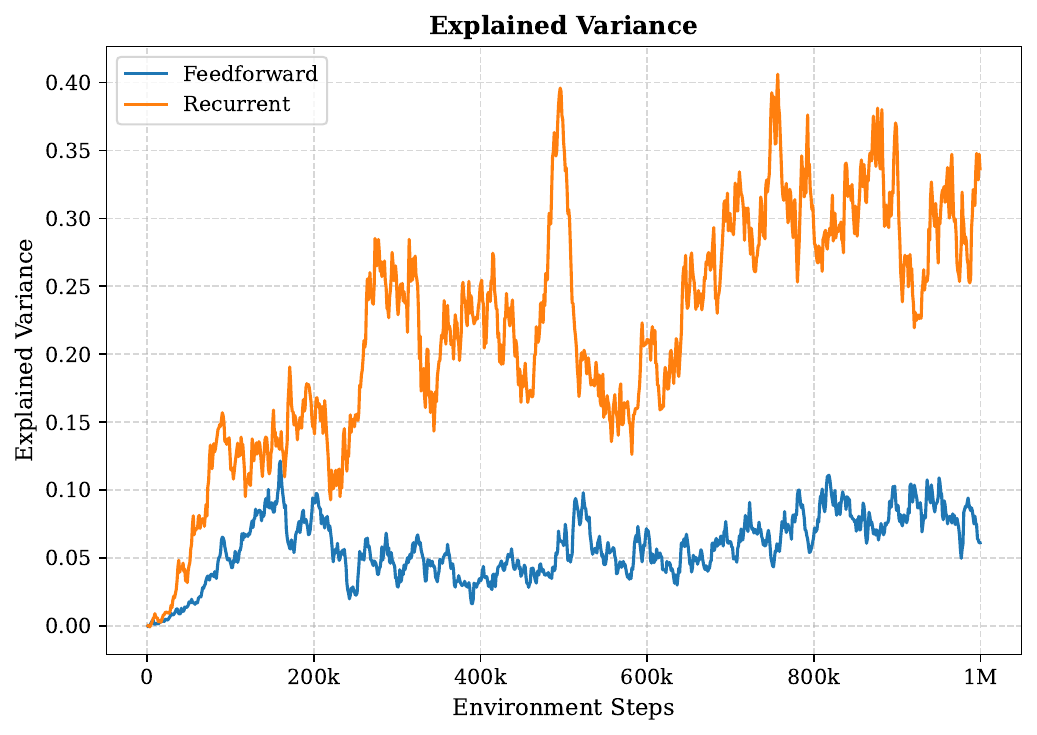}
    \caption{Explained variance of the value function.}
    \label{figure_explained_variance}
    \vspace{-0.5cm}
\end{figure}

\paragraph{Impact of Temporal Memory} We next compared feedforward and recurrent agents. Recurrence improved success rate for both algorithm families. More importantly, the recurrent PPO agent achieved substantially higher explained variance than the feedforward variant, as shown in Figure~\ref{figure_explained_variance}. This indicates that temporal memory improves internal state estimation and yields more stable value learning under ambiguous RF observations.

\begin{figure}[!t]
    \centering
    \begin{minipage}[t]{0.493\linewidth}
        \centering
        \includegraphics[trim=20 20 18 24, clip, width=1.0\linewidth]{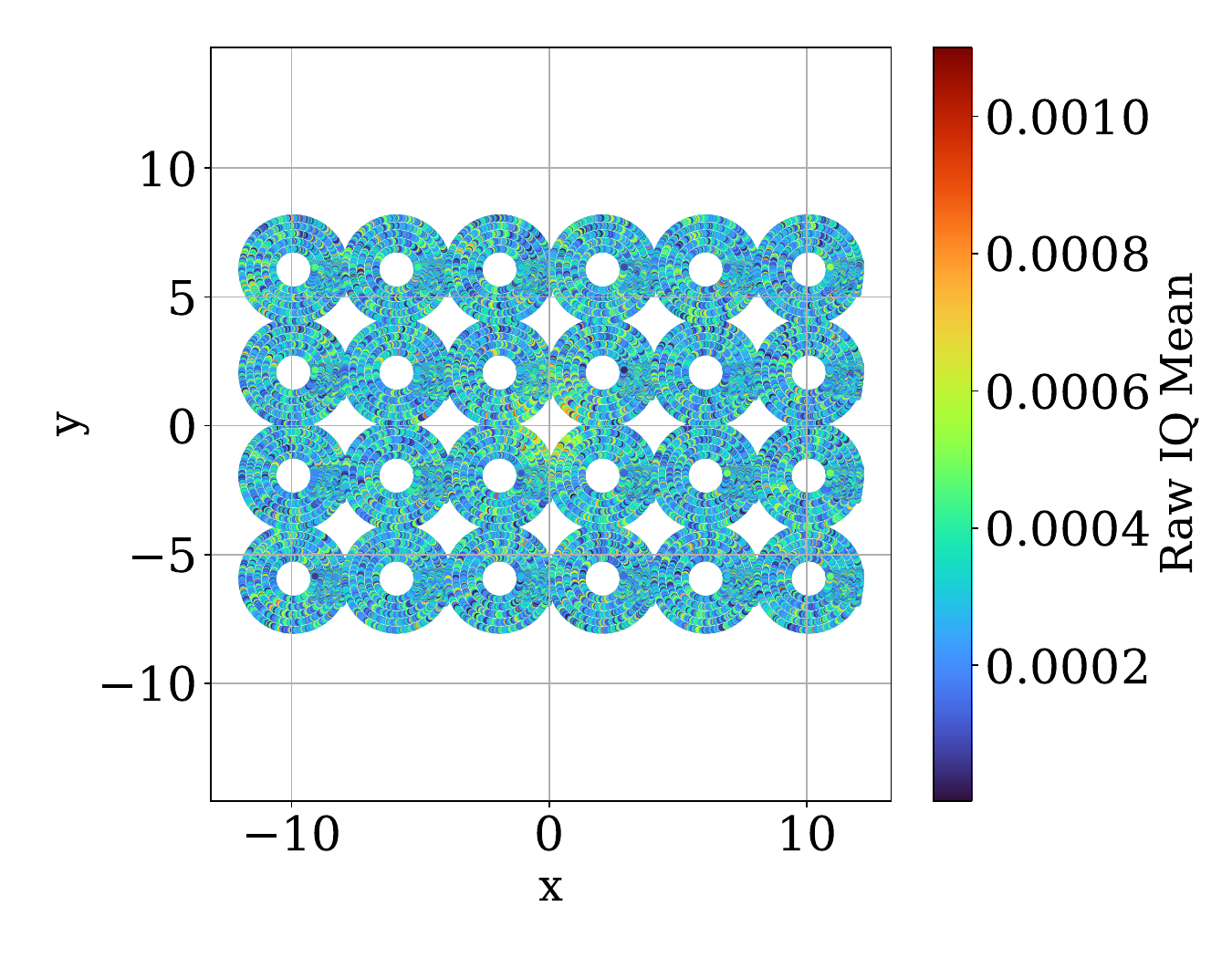}
        \vspace{-0.35cm}
    \end{minipage}
    \hfill
    \begin{minipage}[t]{0.493\linewidth}
        \centering
        \includegraphics[trim=20 20 18 24, clip, width=1.0\linewidth]{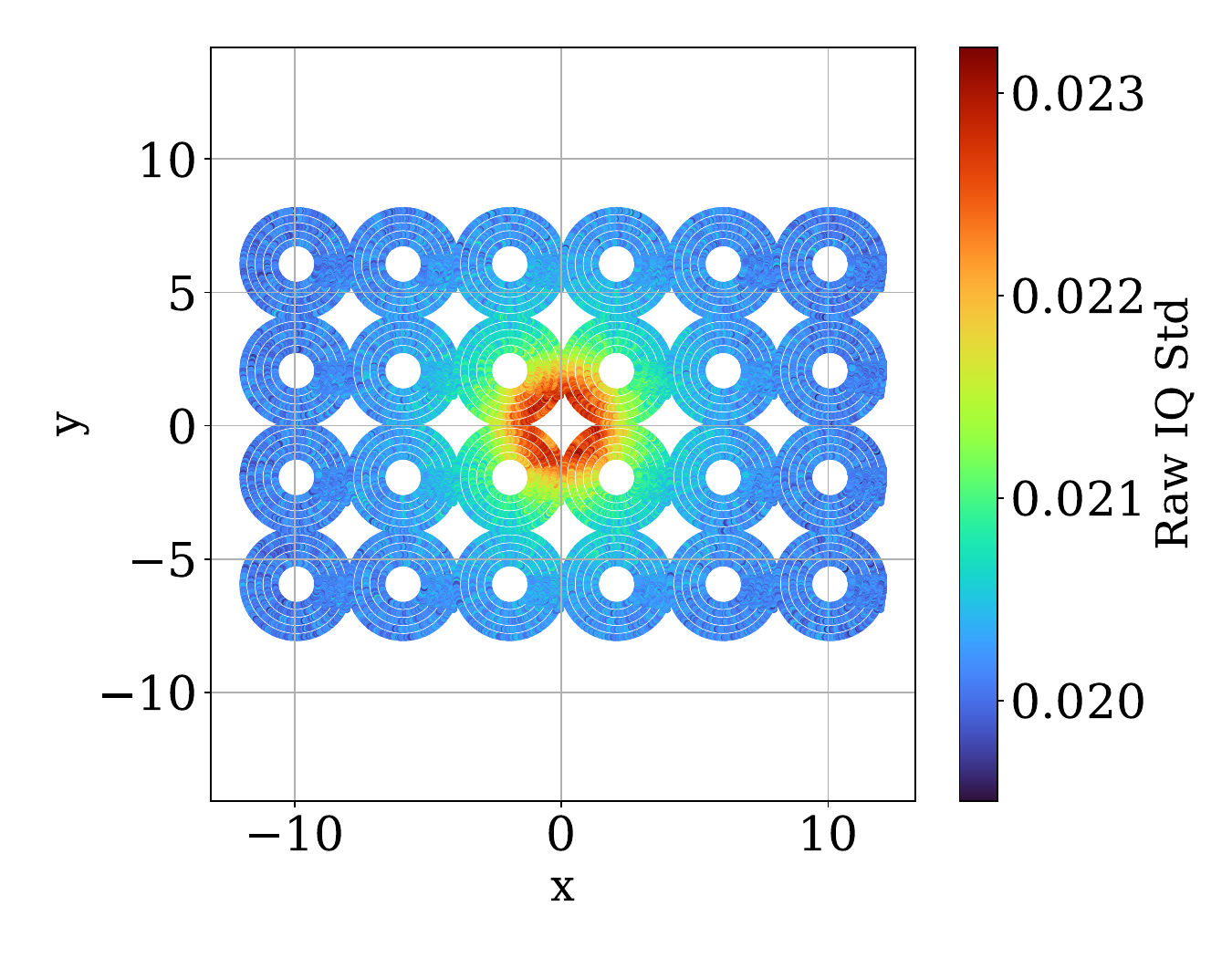}
    \end{minipage}
    \hfill
    \begin{minipage}[t]{0.493\linewidth}
        \centering
        \includegraphics[trim=20 20 18 24, clip, width=1.0\linewidth]{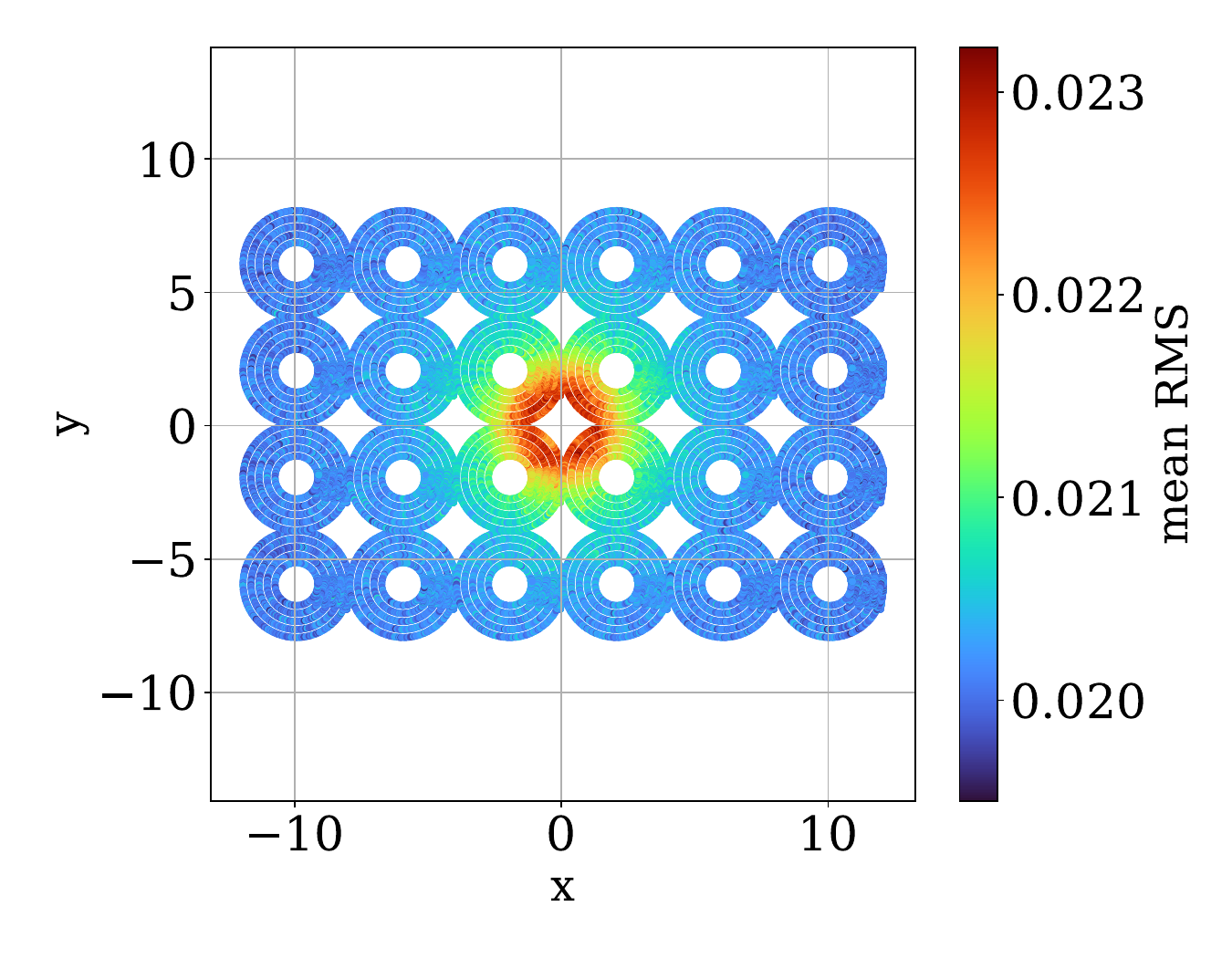}
    \end{minipage}
    \hfill
    \begin{minipage}[t]{0.493\linewidth}
        \centering
        \includegraphics[trim=20 20 18 24, clip, width=1.0\linewidth]{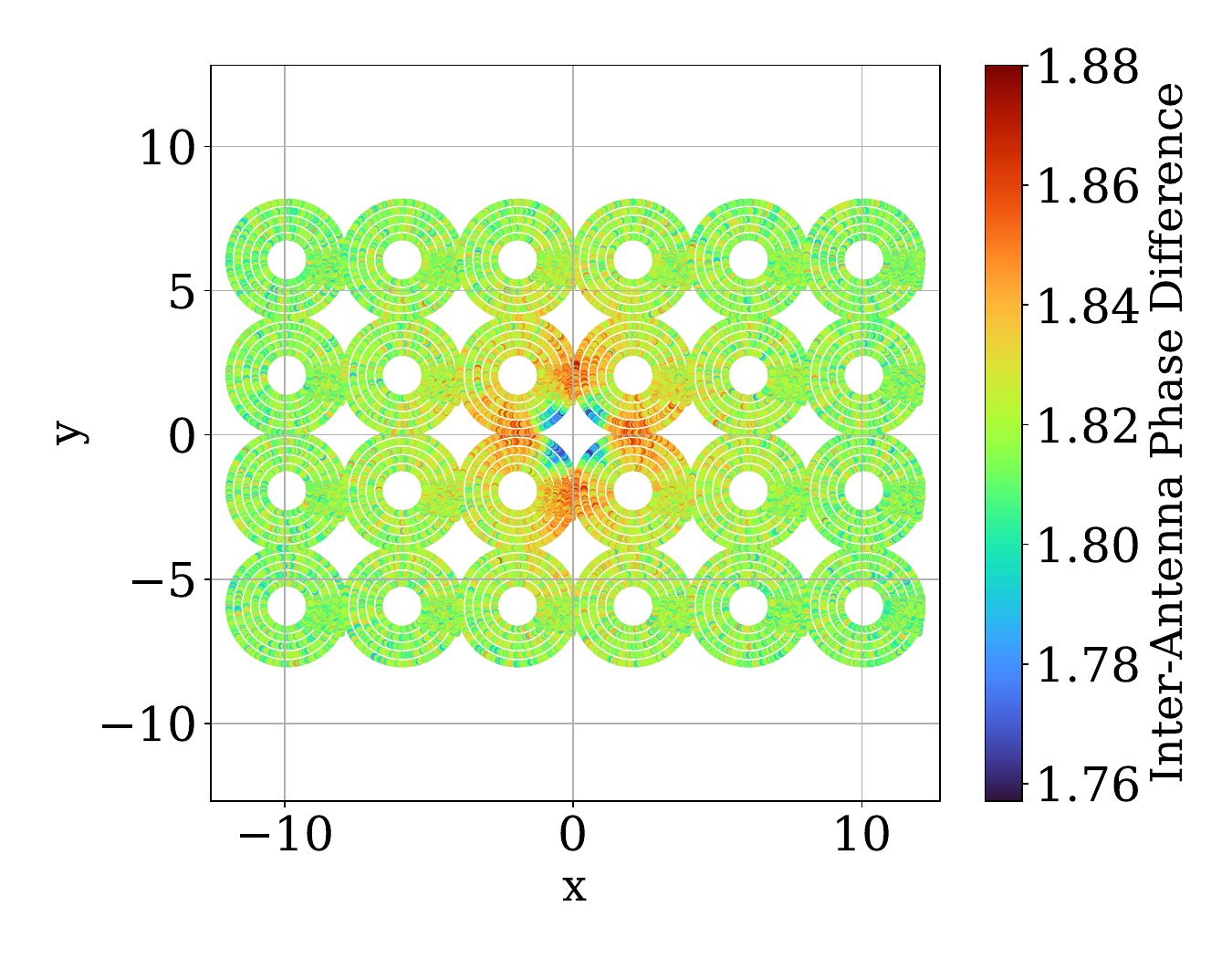}
    \end{minipage}
    \caption{Feature importance for different signal representations.}
    \label{figure_feature_importance}
\end{figure}

\begin{figure}[!t]
    \centering
    \begin{minipage}[t]{0.493\linewidth}
        \centering
        \includegraphics[trim=24 24 24 24, clip, width=1.0\linewidth]{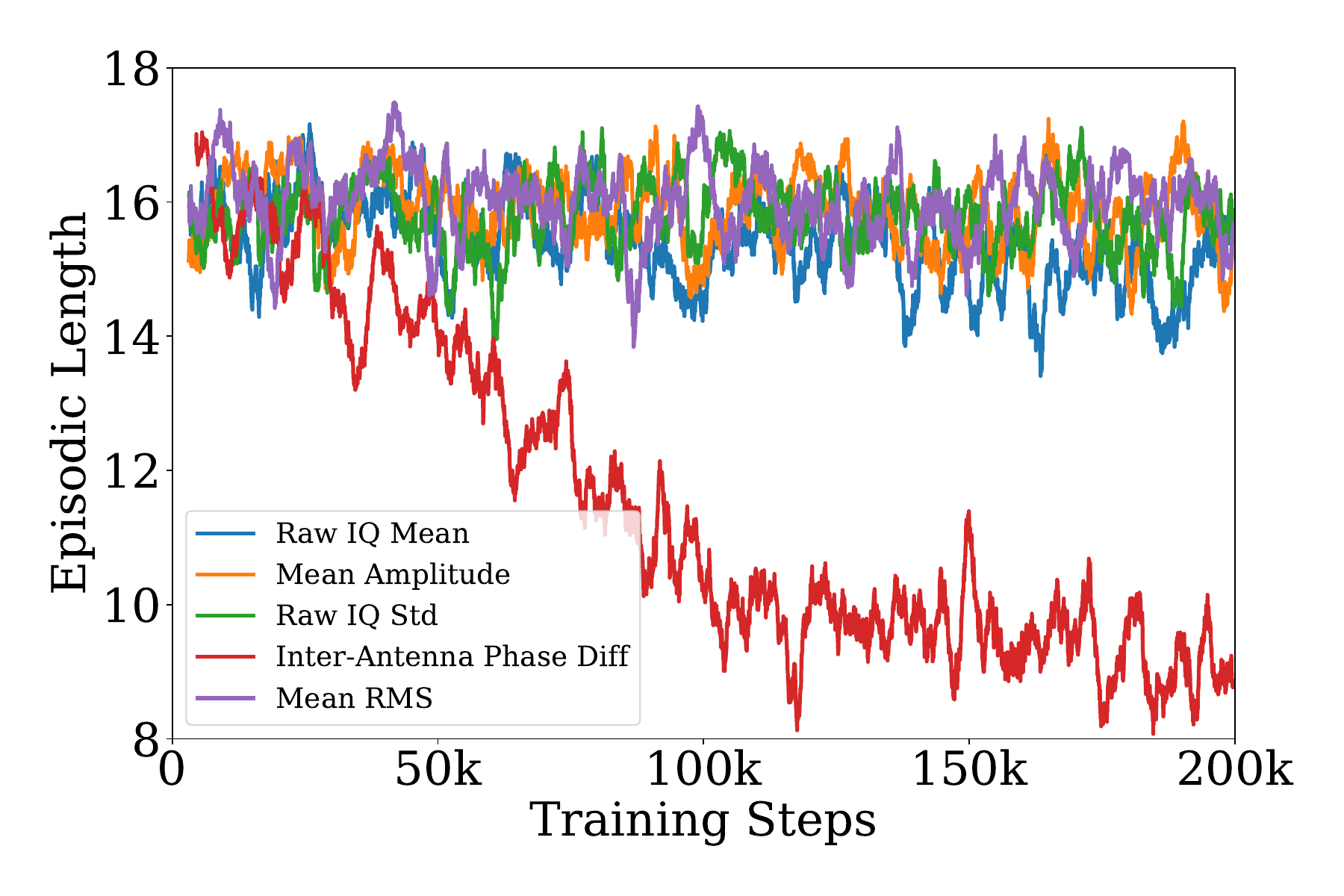}
    \end{minipage}
    \hfill
    \begin{minipage}[t]{0.493\linewidth}
        \centering
        \includegraphics[trim=24 24 24 24, clip, width=1.0\linewidth]{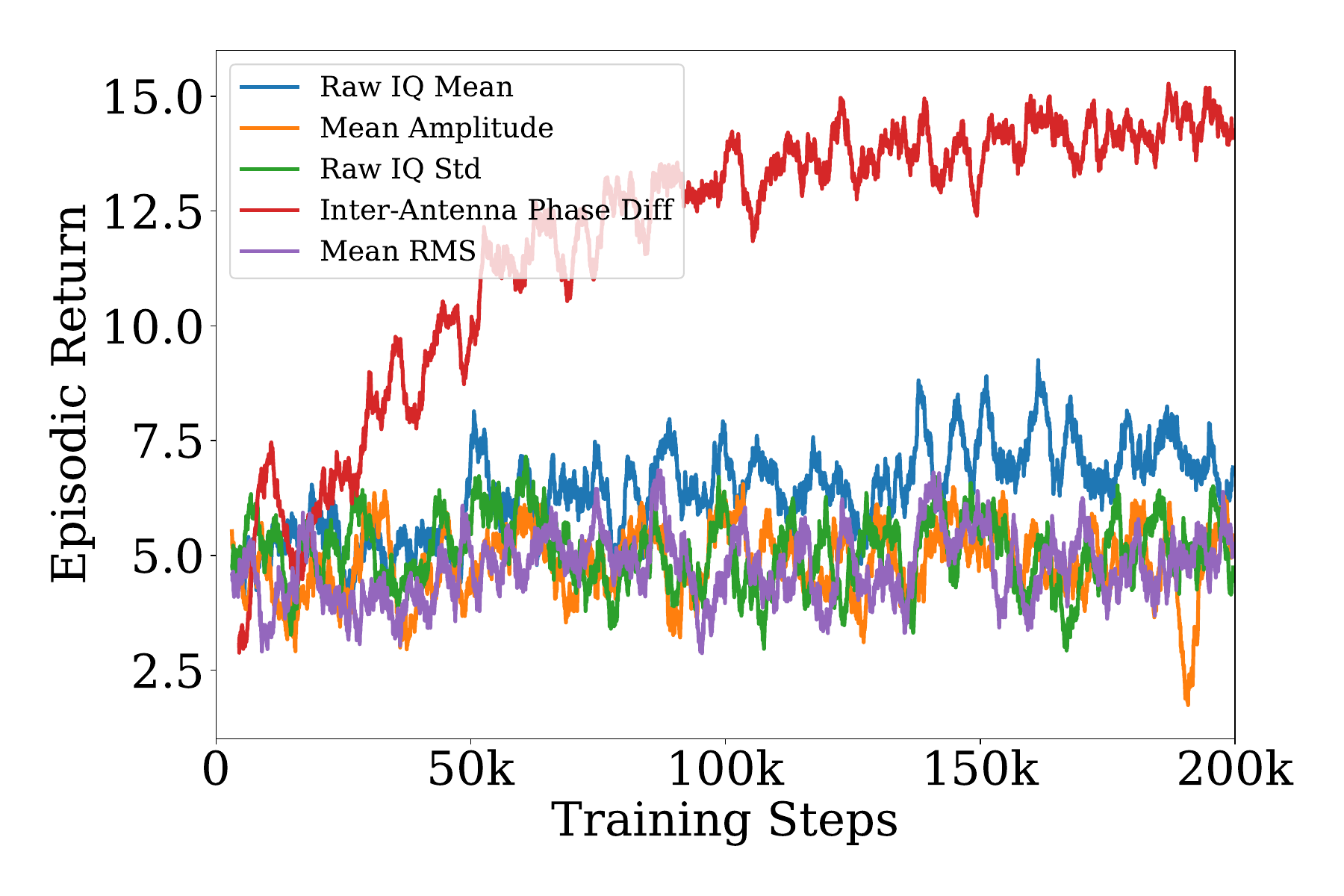}
    \end{minipage}
    \caption{Comparison of episodic length and return values for different input signal representation inputs.}
    \label{figure_sionna_evaluation}
\end{figure}

\paragraph{Feature Selection} Figure~\ref{figure_feature_importance} illustrates the spatial distributions of several signal representations, including the mean, standard deviation, RMS, and inter-antenna phase differences. The mean, standard deviation, and RMS features exhibit similar patterns, with elevated responses concentrated near the center and only limited variation across the remaining cells. By contrast, the inter-antenna phase difference shows a more distinctive and spatially structured distribution, indicating that it provides the most discriminative information for localization. Consistent with this observation, the reinforcement learning agent achieves its best performance under the PPO framework when the phase differences between IQ samples are used as input features. In addition, the phase difference representation leads to more stable learning behavior and higher localization accuracy (see Figure~\ref{figure_sionna_evaluation}). Unlike the other feature types, phase differences remain consistently distinguishable across different cells of the environment, whereas the remaining representations exhibit overlapping and less discriminative patterns that hinder reliable spatial differentiation. The consistency of the phase information therefore provides a stronger and more informative signal, enabling the agent to infer spatial relationships more effectively and to improve its decision-making process.

\section{Conclusion}
\label{label_conclusion}

We formulated signal emitter localization as an active sensing problem and presented an RL framework that learns source-seeking behavior directly from multi-antenna IQ observations. The results show that RL is a viable approach for this task, while recurrent policies improve stability under partial observability and meta-learning enables positive transfer to unseen propagation environments. The findings highlight the potential of adaptive RL for emitter localization.
\blfootnote{\textbf{Acknowledgments.} This work has been carried out within the PaiL project, funding code 50NP2506, sponsored by the German Federal Ministry for Transport (BMV) and supported by the German Space Agency at DLR, the Bundesnetzagentur (BNetzA), and the Federal Agency for Cartography and Geodesy (BKG).}

\bibliography{IPIN2026}
\bibliographystyle{IEEEtran}

\end{document}